# The Endocranial Cast of *Khirtharia* (Artiodactyla, Raoellidae) Provides New Insights into the Earliest Evolution of the Cetacean Brain

Mohd Waqas a, b  Thierry Smith c  Rajendra Rana b  Maeva J. Orliac a a Institut des Sciences de l'Evolution de Montpellier, Université de Montpellier, Montpellier, France;
b Department of Geology, HNB Garhwal University, Srinagar, India;
c Directorate Earth and History of Life, Royal Belgian Institute of Natural Sciences, Brussels, Belgium

Keywords: Cetacea · Brain · Middle Eocene · India

**Abstract**

Introduction: Raoellidae are small artiodactyls retrieved from the middle Eocene of Asia (ca. -47 Ma) and closely related to stem Cetacea. Morphological observations of their endocranial structures allow for outlining some of the early steps of the evolutionary history of the cetacean brain. The external features of the brain and associated sinuses of Raoellidae are so far only documented by the virtual reconstruction of the endocast based on specimens of the species *Indohyus indirae*. These specimens are however too deformed to fully access the external morphology, surface area, and volume measurements of the brain. Methods: We bring here new elements to the picture of the raoellid brain by an investigation of the internal structures of an exceptionally well-preserved cranium collected from the Kalakot area (Jammu and Kashmir, India) referred to the species *Khirtharia inflata*. Micro-CT scan investigation and virtual reconstruction of the endocast and associated sinuses of this specimen provide crucial additional data about the morphological diversity within Raoellidae as well as reliable linear, surfaces, and volumes measurements, allowing for quantitative studies. Results: We show that, like *I. indirae*, the brain of *K. inflata* exhibits a mosaic of features observed in earliest artiodactyls: a small neocortex with simple folding pattern, widely exposed midbrain, and relatively long cerebellum. But, like *Indohyus*, the brain of *Khirtharia* shows unique derived characters also observed in stem cetaceans: narrow elongated olfactory bulbs and peduncles, posterior location of the braincase in the cranium, and complex network of blood vessels around the cerebellum. The volume of the brain relative to body mass of *K. inflata* is markedly small when compared to other early artiodactyls. Conclusion: We show here that cetaceans that nowadays have the second biggest brain after humans derive from a group of animals that had a lower-than-average expected brain size. This is probably a side effect of the adaptation to aquatic life. Conversely, this very small brain size relative to body mass might be another line of evidence supporting the aquatic habits in raoellids.

## Introduction

Within mammals, the cetacean brain stands as an exception; it is second only to humans in terms of size as scaled for body allometry [1, 2]. It also shows a drastic reduction or a lack of olfactory bulbs (microsmatic mammals; e.g., [3–7]), a strong telencephalic flexure [8], and a high complexity of the neocortex with huge cerebral hemispheres (e.g., [8–11]) and special cortical characteristics (i.e., simple cortical organization [12–15]). In relation with underwater hearing, cetaceans also show an extraordinary development of the acoustic system as reflected by the huge size of the caudal colliculi of the midbrain (which receives input from the ears [3, 16]). The brain is partially surrounded by a dense network of arteries and veins called the rete mirabile, "wonderful net" [17, 18], present in artiodactyls but less expanded, and that enables rapid heat exchange and cooling of the arterial blood destined for the brain. The earliest evolutionary history of the cetacean brain is only partly documented by early middle Eocene (ca. −50 to −45 Ma) extinct representatives of the group, pakicetids [7, 19], protocetids [20], and remingtonocetids [7, 21]. The endocast morphology of these extinct families indicates modifications of the olfactory tract (elongation of olfactory peduncles and bulbs [6, 7, 21], and the presence of wide venous sinuses lateral to the brain (associated with the rete mirabile; *Indocetus ramani* [20]; *Remingtonocetus* [21]). However, only few characters are accessible on these fossils, and general characteristics such as volumes and proportions of the various components of the brain remain difficult to assess. Indeed, besides partial preservation, the morphology of the cerebrum and cerebellum is obscured by sinuses and blood vessels (part of retia mirabilia) that make it impossible to accurately determine the volume of the brain based on the intracranial volume in these early cetaceans.

The family Raoellidae [22] became central to the question of cetacean evolution and land to water transition after the work of Thewissen et al. [23] that retrieved raoellids as the closest relatives to stem Cetacea, a result subsequently supported by other analyses (e.g., [24, 25]; Figure 1). Raoellids are mostly documented from middle Eocene deposits of south Asia, coeval with early cetaceans and found in the same fossil yielding localities [23, 26, 27]. The external features of the brain and associated sinuses of Raoellidae are known from the virtual reconstruction of the endocranial cast of two specimens of *Indohyus indirae* from India [28]. This taxon already shows some of the cetacean brain features: an elongation of the olfactory tract and intraosseous blood sinuses above the cerebellum that might represent the initial development of the caudal venous rete mirabile. However, the specimens described by Orliac and Thewissen [28] are unfortunately badly deformed, making impossible to fully access the external morphology of the brain and the surface area and volume measurements. We provide here additional information to the picture of the raoellid brain by an investigation of the internal structures of an exceptionally well-preserved cranium (GU/RJ/297) collected from the fossil locality East Aiji-2 Rajouri District, Jammu and Kashmir [29] and referred to *Khirtharia inflata* [30]. Micro-CT scan investigation and virtual reconstruction of the endocranial cast and associated sinuses of the cranium provide crucial additional data about the morphological diversity within Raoellidae, as well as reliable linear surfaces and volumes measurements, allowing for quantitative studies. The exceptional specimen of *Khirtharia* studied here allows going further into the comparisons of raoellids with other Eocene artiodactyls and provides the first tangible quantitative data to discuss the earliest steps of cetacean brain history.

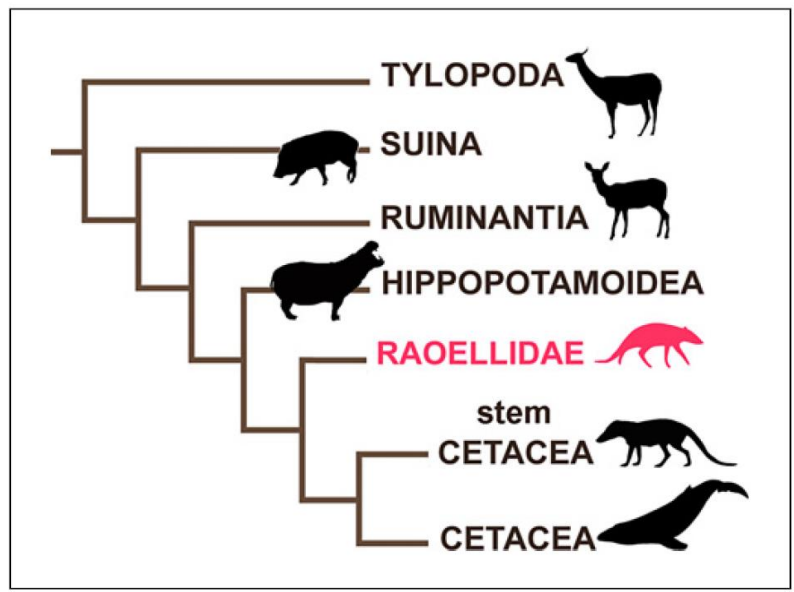

**Fig. 1**. Phylogenetic position of Raoellidae within Artiodactyla based on [24].

**Materials and Methods**

Our present work is based on the cranium of *K. inflata*, GU/RJ/197 described by Waqas et al. [30] and Orliac et al. [31] from the middle Eocene locality of Aiji 2, belonging to the Subathu Group of the Kalakot region in the northwest Himalaya of Rajouri district, Jammu and Kashmir [29, 32]. The specimen represents a fully grown adult with completely erupted M3s and little dental attrition. The specimen is housed in the collections of the Garhwal University, Paleontology lab. The 3D data acquisition was performed at the µ-CT scanner facility of the Montpellier Rio Imaging platform (MRI) at the University of Montpellier, using a RX Solutions EasyTom 150 a-CT scanner. The voxel size is of 69.45 µm. Segmentation and measurements were performed using Avizo ® 9.3 (Thermo Fisher Scientific-FEI). Segmentation was performed manually slice by slice using the pencil segmentation tool. Cranium, endocast, and sinuses were segmented separately on different label fields. Measurements generally follow the protocol of Macrini [33]; telencephalic flexure was measured as indicated in the online supplementary information (online suppl. Fig. S1; for all online suppl. material, see https://doi.org/10.1159/000542574). The cerebrum and neopallium surfaces were measured using the tag tool of MorphoDig [34]. The nerves casts were not included in the surface of the cerebrum. The 3D models of the specimens described in this work are available online for visualization and download on the platform MorphoMuseuM ([35] models id M3#1608, M3#1609).

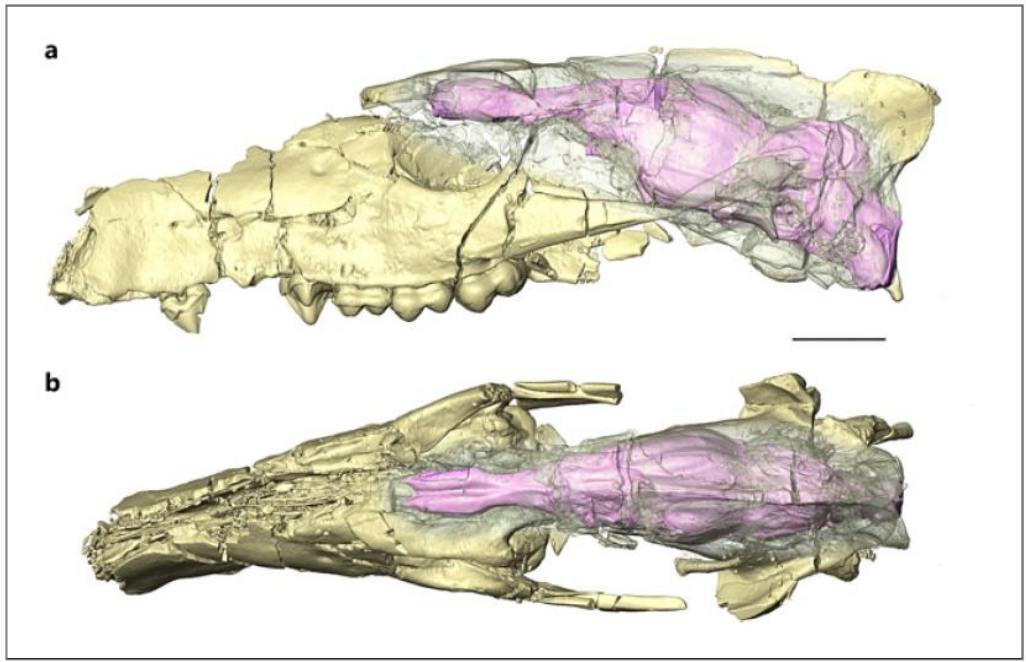

**Fig. 2**. Cranium of *Khirtharia inflata* (GU/RJ/197) showing the location of the braincase (in pink) in lateral (a) and dorsal (b) views. Scale bar = 10 mm.

**Results**

Preservation and General Shape of the Specimen

The cranium GU/RJ/197 underwent only slight lateral compression, and the deriving endocast gives access to the various components of the brain in their original arrangement as well as to good estimates of the width and height of the brain case (Fig. 2). Damages on the ethmoidal, orbital, and temporal areas prevent from accessing to the ventral surface of the olfactory bulbs and on the neocortex surface on the right aspect of the endocast. The brain is elongated and looks stretched especially in the olfactory tract region which length sets the cerebrum back posterior to the orbital region (Fig. 2b). The olfactory bulbs and the cerebrum are the highest points of the brain in lateral view when the palate is horizontal. There is a slight telencephalic flexure of 165°. The cerebrum and the cerebellum are of comparable height, the former having the greatest width. The total volume of the braincase is 5,582 mm$_3$. Measurements of the endocranial parts are listed in Table 1.

Rhinencephalon

The new specimen allows nearly complete examination of the whole rhinencephalon. The cribriform plate is only very partly preserved and most of it is broken away. Yet, the shape of the olfactory bulbs, housed inside the ethmoidal chambers, can be reconstructed. They are slender and elongated and extend over the length of the orbit (Fig. 2). The left and right chambers are united over most of their length. The olfactory bulbs are separated from the cerebral hemispheres by a clear circular fissure (Fig. 3). The whole olfactory tract accounts for around 25% of the total length of the endocranial cast in dorsal view and has a volume of 311mm$_3$ which represents 5.6% of the total braincase volume. There is not much preservation in the region of the olfactory tubercles, but there does not seem to be an important swelling in their place. In lateral view, the location of the posterior rhinal fissure indicates that the piriform lobes were relatively large in comparison to the size of the neocortex. The piriform lobes bear a clear delineation of a sulcus, lying ventral to the orbitotemporal canal.

|  | *Khirtharia inflata* | *Indohyus indirae* | |
| --- | --- | --- | --- |
|  | GU/RJ/197 | RR 207 | RR 601 |
| Endocast maximum length | 64.8 | 74.2 | 73.1 |
| Endocast total volume | 5,582 |  |  |
| Olfactory bulb cast anteroposterior length | 12.1 | 18.0 | 15.6 |
| Olfactory bulb cast maximal width | 6.72 (left)[a] 7.44 (right) | 11.7 | 11.6 |
| Olfactory tract length anterior to cerebrum | 16.2 | 24.9 | 23.3 |
| Olfactory bulbs volume – measured | 311[a] |  |  |
| Olfactory bulbs volume – estimated | 486 |  |  |
| Cerebrum cast anteroposterior length | 27.38 | 32.4 | 28.8 |
| Cerebrum cast maximal width | 15.83 |  | 27.4[a] |
| Neocortex surface | 262 |  |  |
| Midbrain maximal exposure (estimated) | 1.23 | 3.6 | 4.0 |
| Cerebellum cast anteroposterior length | 14.90 | 11.6[a] | 12.4[a] |
| Cerebellum cast maximal width | 13.21 |  |  |
| Vermis width (estimated) | 3.9 |  |  |

[a] Marks underestimates values due to breakage or deformations.

**Table 1**. Measurements of the endocranial casts of *Khirtharia inflata* and *Indohyus indirae* [28] in mm (linear), mm2 (surface area), and mm3 (volumes)

Neopallium

The neopallium is delimited ventrally by the rhinal fissure, and the two cerebral hemispheres are separated medially by a salient dorsal sagittal sinus. The entire cerebral hemispheres are posterior to the orbit (Fig. 2b). The anterior most limit of the neopallium is almost continuous with the olfactory peduncles and difficult to identify. Yet, on the left side, a swelling marks the anterior margin and what is here tentatively identified as the presylvia (online suppl. Fig. S2). This interpretation implies that the neopallium is stretched anteriorly, just like the olfactory tract. Posteriorly, the rhinal fissure runs in a rather dorsal position, and the piriform lobe is large. The orbitotemporal canal traverses the piriform lobe well below the level of the rhinal fissure, highlighting the decoupling between the location of both structures in early artiodactyls (the orbitotemporal canal is considered as hallmark of the rhinal fissure in other mammal groups such as rodents and primates, e.g., [36, 37]). The neopallium has a surface area of 262 mm2, which represents 47.5% of the surface of the cerebrum. The neocortex is gyrencephalic and bears three sulci. The suprasylvia and the lateral sulci delimit a narrow almond-shaped gyrus 3; they merge anteriorly in a deep and extended depression known as the coronal sulcus (Fig. 3a, c).

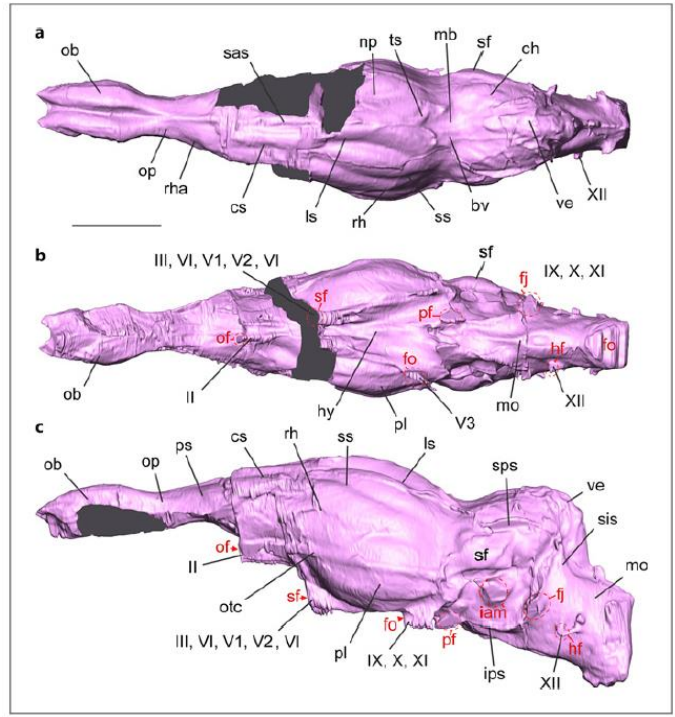

**Fig. 3**. Labelled endocranial cast of *Khirtharia inflata* (GU/RJ/197) illustrated in dorsal (a), ventral (b), left lateral (c) views. Casts of foramina area labelled in red, broken surfaces appear in dark grey. bv, blood vessel; ch, cerebellar hemisphere; cs, coronal sulcus; fj, foramen jugulare; fo, foramen ovale; hf, hypoglossal foramen, hy, hypophysis; iam, internal acoustic meatus; ips, inferior petrosal sinus; ls, lateral sulcus; mb, midbrain; mo, medulla oblongata; np, neopallium; ob, olfactory bulb; of, optic foramen; op, olfactory peduncle; otc, orbitotemporal canal; pf, piriform fenestra; pl, piriform lobe; ps, presylvia; rh, rhinal fissure; rha, anterior rhinal fissure; sas, dorsal sagittal sinus; sf, subarcuate fossa; sis, sigmoid sinus; sp, sphenorbital fissure; sps, superior petrosal sinus; ss, suprasylvia; ve, vermis. Impressions of foramina are indicated in red dashed lines. II to XII refer to cranial nerves. Scale bar = 10 mm.

### Midbrain and Cerebellum

The cerebrum is not extended posteriorly and its posterior margin does not abut the cerebellum. However, the tectum of the midbrain is not exposed dorsally and is covered by blood vessels. Only a shallow depression lies between the posterior margin of the neopallium and the cerebellum, anterior to the vermis, (Fig. 3c). The detailed external morphology of the cerebellum is blurred by a network of sinuses and blood vessels that hampers determining precisely the limits of the vermis and the extension of the cerebellar hemispheres on the dorsal surface of the endocast. In dorsal view, the cast of the subarcuate fossa protrudes laterally, but there is no clear paraflocculus. A small swelling is present medial to the subarcuate fossa cast. The width of the vermis is accessible in the posterior part of the cerebrum, it appears to be narrow, but this might be partly due to compression of the specimen. Laterally, the imprint of the petrosal bone is deep. The swelling of the subarcuate fossa overhangs the cast of the internal auditory meatus with the vestibulocochlear nerve exit. The vermis is relatively high, and its sulci might be visible in its posteriormost portion notably the secondary fissure (Fig. 3a, c).

### Nerves and Sinuses

On the ventral side of the endocast, the anterior most portion of the olfactory bulbs, the cribriform plate isunfortunately not preserved, and the foramina for the olfactory nerve fascicles (cranial nerve I [CN I]) cannot be reconstructed (Fig. 3b). Slightly posterior to the circular fissure, the exit of the optic nerves (CN II) is visible as well as the cast of the optic nerve pathway (Fig. 3c), visible posteriorly to the level of the sphenorbital fissure (transmitting vessels and cranial nerves III to VI, except for CN V.3). Like in *Indohyus* specimens [28], two separate grooves lead to this fissure: the medial one likely corresponds to the pathway of the oculomotor nerve (CN III), while the lateralmost might correspond to a separate pathway for CN V1 and V2. The pituitary fossa is small and shallow and lies in an anterior position, at the stem of the sphenorbital canal. The foramen ovale, transmitting CN V3, lies far posterior to the sphenorbital fissure and lateral to it. The casts of the facial (CN VII) and vestibulocochlear (CN VIII) nerves are prominent, protruding on the lateral aspect of the endocast. The slit of the basicapsular fissure cast connects the cast of the piriform fenestra to the jugular foramen (CN IX, X, XI). Posterior to it, the medulla oblongata exhibits the cast of the hypoglossal foramen, where CN XII leaves the skull.

The sagittal sinus is prominent dorsally and extends from the circular fissure to the posterior part of the neopallium to the level of the transverse sinus, lining the posterior margin of the cerebral hemispheres (Fig. 3a). The detail of the dorsal surface of the cerebellum is blurred by the presence of a network of blood vessels and by the cast of the superior petrosal sinus. In lateral view, the casts of the sigmoid and inferior petrosal sinuses are also visible (Fig. 3b, c). Like *Indohyus*, *Khirtharia* shows intraosseous sinuses, illustrated in Figures 4–5. A large temporal sinus [38] (pathway for the capsuloparietal emissary vein [39]) is present laterodorsally and anteriorly to the petrosal bone. Its anterior part opens dorsally onto a wide temporal foramen (temporal foramina = foramina for ramitemporales; see [40]) located close to the nuchal crest; a second smaller foramen is present on the left side (Fig. 4d, 5). Posteriorly, the temporal sinus cavity connects to the sigmoid sinus pathway (corresponding to the posterior distributary branch of the transverse sinus [41]). It opens ventrally, posterior to the glenoid surface into the postglenoid foramen, which gives pathway to the capsuloparietal emissary vein, [42, 43]. There is no trace of secondary post-glenoid foramen [28]. Like in *Indohyus*, the left and right temporal sinuses are connected by a dorsal confluence of sinuses, at the level of the midbrain. This network of small cavities resembles in part the spongious bone, but spaces are interconnected and open onto the cranial cavity in several points (Fig. 4b, d). One of the small canals connects the posterior part of the sagittal sinus and two thin canals run on the left and right laterodorsal sides, into the parietal bone (Fig. 4a, 5a, b). Posteriorly, a larger canal opens on the lateral aspect of the cranium, close to the sagittal crest (Fig. 5b, c). Beneath the intraosseous network, the cast of the endocranial cavity at the midbrain level shows two longitudinal ridges corresponding to undetermined blood vessel pathways (Fig. 3a). Posteriorly, the intraosseous sinus lying between the parietal and occipital bones and within the occipital bone is very reduced (posterior occipital sinus [pos], Fig. 5; ?occipital sinus of Butler [44]) and seems to mostly locate into the cranial cavity. Like in *Indohyus*, it opens into the cranial cavity at the level of the sigmoid sinus (Fig. 5b), and outside the cranium, on the occipital aspect of the skull, at the level of the mastoid foramen (Fig. 5c). The mastoid foramen might have given pathway to the occipital emissary vein [39, 45] and/or to the vena diploetica magna [43]. Small channels, exiting dorsal to mastoid foramen, might have given pathway for emissaries of the diploic vein. The pathway for the condyloid vein is also visible, lateral to the foramen magnum.

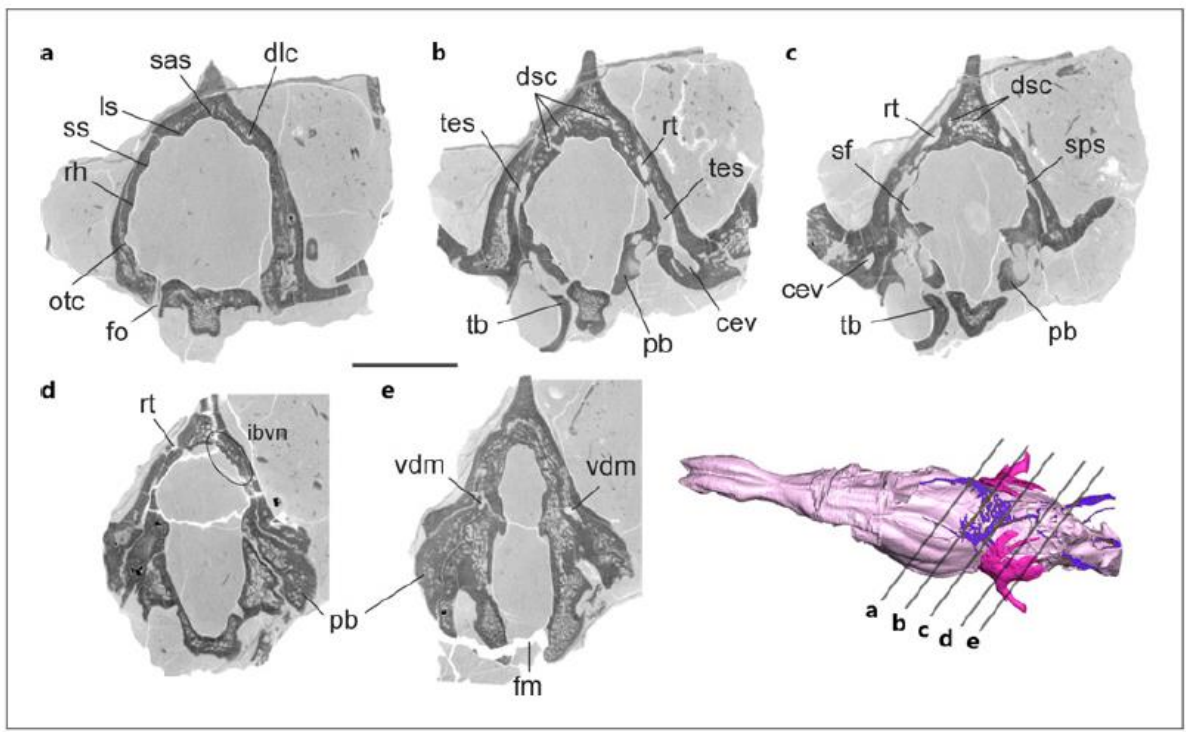

**Fig. 4**. a–e Labelled CT slices through the cranium *Khirtharia inflata* (GU/RJ/197); the location of slices is indicated on the endocast at the bottom right; pink, cast of the main cavity of the braincase; violet, intraosseous sinuses; magenta, intraosseous sinuses related to the temporal sinus. cev, canal for capsuloparietal emissary vein; dlc, dorsolateral canal; dsc, dorsal confluence of sinuses; fm, foramen magnum; fo, foramen ovale; ls, lateral sulcus; ibvn, intracranial blood vessel network; otc, orbitotemporal canal; pb, petrosal bone; rh, bone swelling marking the rhinal fissure; rt, canal for ramus temporal; sas, sagittal sinus; sf, subarcuate fossa; sps, superior petrosal sinus; ss, bone swelling marking the suprasylvia; tb, tympanic bulla; tes, temporal sinus; vdm, vein diploetica magna. Scale bars =10 mm.

## Brain Size, Encephalization Quotient

The most widely used measure of encephalization is the encephalization quotient (EQ [46]), defined as the quotient of the observed mass of the brain over the expected mass of the brain given a body mass. The brain volume of *Khirtharia* equals 5.58 cm3, which corresponds to a brain mass of 5.78 g (using the brain tissue density of 1.036 g/cm3 [47]). The body mass of *K. inflata* can be estimated using dental dimension based on correlations established on modern mammals [48]. Using lower molar row length (26.4mmbased on GU/RJ/179 [30]), the body mass calculated for *K. inflata* equals 6.9 kg; if we use the M2 length (8.29 mm based on GU/RJ/179 [30]: tab3), it equals 7.4 kg, and if we use the M2 area (0.80 based on mean value of M2 specimens in [30]: tab2), it equals 13.3 kg (see online suppl. information for details of body mass estimates calculations). The body mass estimate for *Khirtharia* is superior when upper dentition is considered. Upper molars are indeed particularly wide compared to cranium size in raoellids, and we therefore prefer to rely on lower molar dimensions. We retain here the body mass estimates of 6.9 kg based on lower molar row length as it provides the best correlation for modern ungulates [48]. The duality between cetaceans and noncetacean artiodactyls in terms of brain size has been extensively described in the literature and modern representatives of both groups present different brain mass/body mass allometric relationships (for review, see [49]). Orliac et al. [49] provide two separate equations, one for non-cetacean artiodactyls, with an expected brain mass = $0.3405(body\ mass)^{0.5603}$, and one for cetacean artiodactyls, with an expected brain mass = $16.0007(body\ mass)^{0.3490}$. Given their status of closest taxon to Cetacea, we calculated the EQ of *K. inflata* using both equations: using the all artiodactyl equation, the EQ equals 0.12 and using the cetacean equation, it equals 0.016. These values are based on a slightly deformed endocast. The deformation, mostly lateral compression, affects the measure of the volume. These EQ values are therefore slightly underestimated.

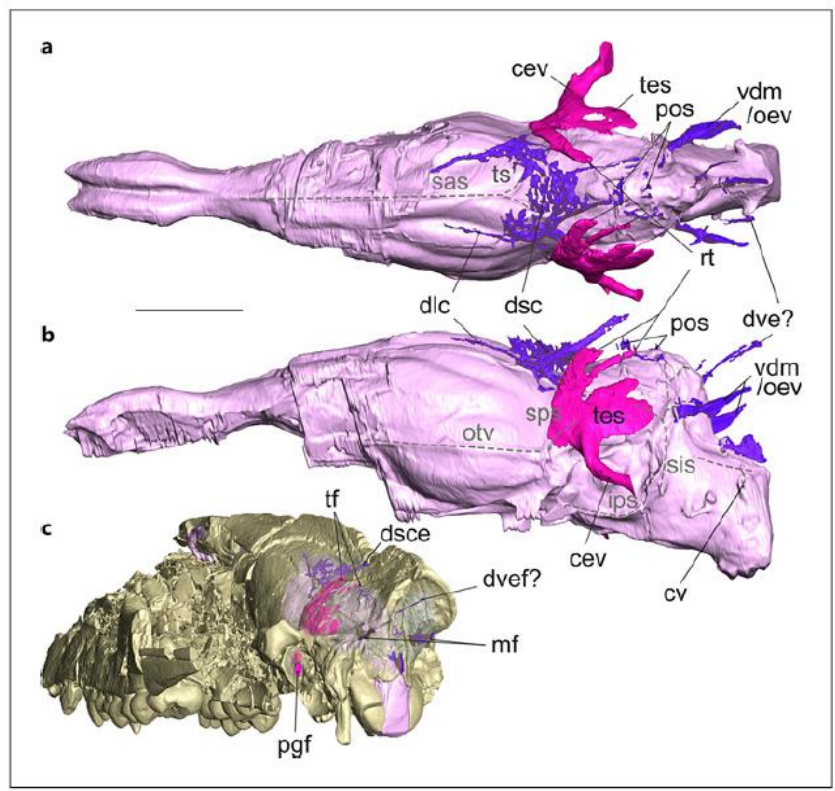

**Fig. 5**. Labelled 3D model of the endocranial cast and associated sinuses of *Khirtharia inflata* (GU/RJ/197) in dorsal (a), lateral (b), posterolateral (c) views; pink, cast of the main cavity of the braincase; violet, intraosseous sinuses; magenta, intraosseous sinuses related to the temporal sinus. cev, canal for capsuloparietal emissary vein; cv, condyloid vein; dlc, dorsolateral canal; dsc, dorsal confluence of sinuses; dsce, dorsal sinus canal exit; dve, diploic vein emissary; dvef, diploic vein emissary foramen; ips, inferior petrosal sinus; mf, mastoid foramen; oev, occipital emissary vein; otv orbitotemporal vein; pgf, post-glenoid foramen; pos posterior occipital sinus; rt, rami temporale; sas, sagittalsinus; sis, sigmoid sinus; sps, superior petrosal sinus; tes, temporal sinus; tf, temporal foramen; vc, vascular canal; vdm, vein diploetica magna. Scale bars = 10 mm.

**Discussion**

Completing Our Knowledge of the Raoellid Brain Morphology

The endocranial morphology of *Khirtharia inflata* is congruent with that observed for *Indohyus indirae* based on the specimens RR 207 and RR 601 described and illustrated by [28]. The specimen described here has undergone only light lateral compression and provides access to the height of the brain cavity and to the morphology of the lateral aspect of the cerebrum, which were inaccessible on the *Indohyus* specimens due to intense deformation (online suppl. Fig. S3). The *Khirtharia* specimen confirms that the neopallium in raoellids is small and the piriform lobe extended, as supposed based on the dorsoventrally crushed crania of *Indohyus*. The anterior extension and elongation of the olfactory bulbs, setting back the braincase posterior to the postorbital rim is also confirmed here. Yet, the good preservation of the anterior most part of the cerebrum indicates that the neopallium extends more anteriorly than previously described on *Indohyus*, and that a presylvia is most probably present. A main difference between the reconstructions of *Indohyus* and *Khirtharia* is the lack of dorsal exposure of the colliculi in the latter. There are two salient structures at the midbrain level, but they clearly reflect the presence of blood vessels instead of colliculi protrusion (Fig. 3a). This questions the actual presence of colliculi exposure in *Indohyus* described by Orliac and Thewissen [28] that finally might correspond to imprints of blood vessels instead. The schematic reconstruction of the venous pattern of the posterior portion of the raoellid cranium proposed by Orliac and Thewissen [28] based on *Indohyus* is congruent with the observations on the *Khirtharia* specimen. However, regarding sinuses and vein pathways, *Khirtharia* shows some differences with what was reconstructed for *Indohyus* (online suppl. Fig. S4). Regarding the temporal sinus and associated structures, there is no trace of secondary post-glenoid foramen, and no tributary to the capsuloparietal emissary vein opening dorsal to the glenoid surface, suggesting variability in the presence of this structure, at least at the familial level. The extension of the temporal sinus is different in *Khirtharia* and *Indohyus*, but this is most probably due to the strong dorsoventral compression of the *Indohyus* specimens, altering the shape of the intracranial space. The sinus overhanging the midbrain also looks slightly different, but the dorsal confluence is present in both taxa. The most striking difference lies in the posterior occipital sinus: the intraosseous course of the latter is much reduced in the *Khirtharia* specimen and seems to be mostly located in the braincase (online suppl. Fig. S4). The presence of intraosseous sinuses in the posterior part of the cranium might be linked to the thickening of the bone observed in *Indohyus* [50] and *Khirtharia* [30] and could be concurrent with aquatic adaptations like suggested in other mammalian groups [51]. Orliac and Thewissen [28] proposed that the intraosseous space dorsal to the cerebellum housing a network of veins and arteries could represent the first steps of an incipient caudal venous rete mirabile. A rete mirabile is a complex of arteries and veins lying very close to each other creating a net. A caudal endocranial arterial and venous rete mirabile is present in modern cetaceans [52–54] and considered present in basilosaurids [43, 52, 53] and protocetids [20]: fig.1A. In the *Khirtharia* specimen described here, the blood vessels network is mostly inside the cranial cavity which would support the link between the dorsal posterior sinus and an incipient caudal rete mirabile. Whether these differences are linked with a different degree of aquatic adaptation remains an open question. The well-preserved endocast GU/RJ/197 allows going further into the comparisons of raoellid with other Eocene artiodactyls. The percentage of cerebral surface covered by the neopallium in *Khirtharia* equals 47.5%. This value is comparable to that of early Eocene artiodactyls such as the North American diacodexeid *Diacodexis ilicis* (42.8%) and homacodontid *Homacodon vagans* (44.3%) or the European dichobunids *Mouillacitherium elegans* (43.2%) and *Dichobune leporina* (44.2%) (online suppl. Table S1). It is however inferior to that of middle and late Eocene European cainotheriids *Caenomeryx filholi* (66.3%), *Anoplotherium* sp.

(61.6%), lying higher in the Artiodactyla tree [55, 56], and inferior to that of the early ruminant *Leptomeryx* (56.6%). *Khirtharia* is therefore close to early diverging artiodactyls in terms of neocortex expansion and is plesiomorphic for this trait. In terms of neocortical complexity, *K. inflata* presents a very simple folding pattern similar to that of *D. ilicis, M. elegans* (Fig. 6a, b), *H. vagans*, and *D. leporina* ([49]: fig.13.4), but with a longer coronal sulcus. A long coronal sulcus is also observed in cainotheriids and in early ruminants like *Leptomeryx* ([49]: fig.13.4). The percentage of the olfactory bulbs volume is 5.6%, which is inferior to that of *D. ilicis* (13.8%), *D. leporina* (7.7%), *Cebochoerus* sp. (7.0%), *Anoplotherium* sp. (7.5%). It is however comparable to that of *M. elegans* (5.8%), *Agriochoerus* sp. (5.0%), *Leptauchenia* sp. (5.5%), and the early ruminant *Leptomeryx* (5.15%). If there is no difference in terms of relative volume of the olfactory bulbs, the morphology of the olfactory chamber with elongated bulbs and peduncles is strikingly different between raoellids and other Eocene artiodactyls (Fig. 6a–d). These differences most probably accompany the deep modification of the cranial architecture in raoellids and do not mean that the olfactory capabilities of *Khirtharia* were affected. It is worth noting that, compared to the other Eocene artiodactyls, there also seems to be no swelling of the olfactory tubercle and a smaller piriform lobe (Fig. 6e–h). Regarding the cerebellum, the external morphology of the latter is obscured by a rather dense network of blood vessels. In terms of proportion, the relative anteroposterior length of the cerebellum of *K. inflata* is similar to that of early artiodactyls *D. ilicis* (Fig. 6a) or *H. vagans* ([49]: fig.13.4), with a length of about half of the cerebrum. Oligocene artiodactyls tend to present a smaller cerebellum length relative to the cerebrum. Just like *Indohyus indirae* [28], the brain of *K. inflata* exhibits a number of features observed in the earliest artiodactyls: a simple neocortical folding pattern, small neocortex expansion, widely exposed midbrain, and relatively long cerebellum. But, on the other hand, the endocast of *Khirtharia*, like that of *Indohyus*, shows unique characters also observed in early cetaceans: narrow elongated olfactory bulbs and peduncles, accompanied by a posterior location of the braincase in the cranium, and complex network of blood vessels. The two known raoellid species present the same general characteristics of the brain. These characters, also observed in extinct Paleogene stem cetaceans, may represent synapomorphies linking Raoellidae to total clade Cetacea.

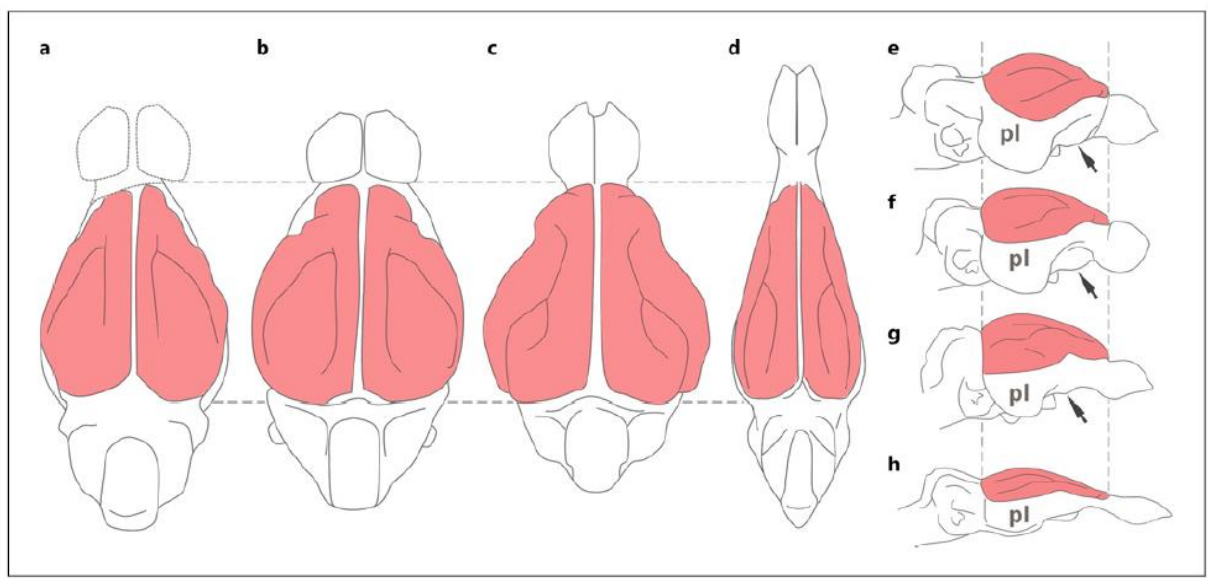

**Fig. 6**. Comparison of the general morphology of the endocast scaled to the size of the neocortex (in red) of *Diacodexis ilicis* [49] (a), *Mouillacitherium elegans* [49] (b), *Leptomeryx* sp. [49] (c), *Khirtharia inflata* (d) (this work, illustrated with deformations) in dorsal (a–d) and right lateral (e–h) views. Arrows point to the olfactory tubercle swelling. pl, piriform lobe. Not to scale

## The Matter of Brain Size

Description of the endocast of *Indohyus* has provided the first insights into the raoellid brain morphology, but it was too deformed to access the volume of the braincase. The specimen of *Khirtharia*, yet slightly deformed, provides here access to the volume of the brain. It has a braincase volume of 5.58 cm3, which corresponds to a brain mass of 5.78 g. The relationship between brain mass and body mass can be simply illustrated by plotting the logarithm of the brain mass against the logarithm of the body mass (Fig. 7). This representation illustrates well the small volume of the brain of *Khirtharia* when compared to other early artiodactyls, even we retrained the lowest body mass estimates. The endocast of *Khirtharia* is laterally compressed, lowering artificially its volume. Yet, even if we correct the volume by a 25% increase, *Khirtharia* remains below other Artiodactyla in the graph. Linear models of the ln (brain mass) relative to the ln (body mass) in extant and extinct species of noncetacean artiodactyls and Cetacea show a difference between the slopes of each group over time, highlighting the fact that the regression equations of each group are different ([49]: fig. 8a).

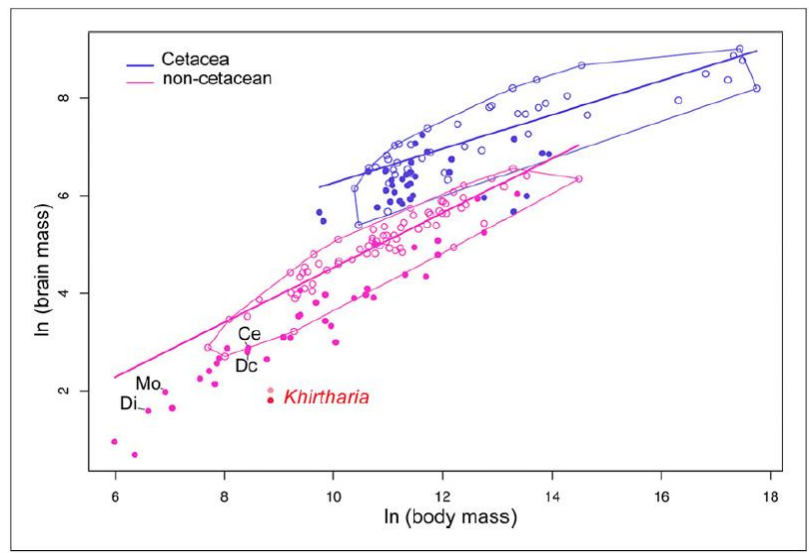

**Fig. 7**. Encephalization within Artiodactyla expressed by an ln brain mass versus ln body mass plot. Empty circles indicate extant taxa; full circles indicate fossil taxa; lines indicate regression lines of extant groups; convex hulls group extant taxa; values for *Khirtharia* appear in red with dark red corresponding to the measures value and light red to an arbitrary 25% increase of the volume. Brain and body data are compiled in [49]. Ce, *Cebochoerus* sp.; Dc, *Dichobune leporina*; Di, *Diacodexis ilicis*; Mo, *Mouillacitherium elegans*.

The brain mass/body mass allometry is a bit more negative in Cetacea and may result from a massive increase of body mass in this clade (see [57]: fig. 1) concurrent with aquatic life and release from the constraints of gravity. Still, modern cetaceans tend to have bigger brains relative to their body size than terrestrial artiodactyls, despite a considerable increase in their body mass. According to Montgomery et al. [57], the EQ of the last common ancestor of cetaceans and hippopotamids, their living sister group, has been estimated at 0.465 using an EQ equation based on a sample of modern mammals established by Jerison [46]. This measure is not directly comparable to the EQ we calculate here using an equation dedicated to artiodactyls. Yet if we apply the EQ equation of [46], i.e., EQ = measured brain mass/($0.12 \times$ Body mass$^{0.67}$), the value for *Khirtharia* is 0.13 instead of 0.12 with the "Artiodactyl equation" of [49]. This value is inferior to the estimates of Montgomery et al. [57] for the cetacean/hippopotamid common ancestor. It is worth noting that the earliest terrestrial artiodactyls of our sample have EQ values close to this estimate (i.e., between 0.45 and 0.58; online suppl. Table S2). Raoellids present a thickening of their skeleton, including long bones [58, 59] and cranium parts [30, 50], interpreted as aquatic adaptation and suggesting that they spent a great deal of their time in water. A heavier skeleton and increased cranium size with elongated snout would have induced increased body mass, compensated by water buoyancy. Yet brain size of raoellids remains comparable to that of their terrestrial counterparts: instead of promoting brain size increased in cetaceans as previously hypothesized (e.g., [8, 60]), adaptation to aquatic lifestyle might have first induced a decoupling between brain size and body mass, shifting the allometric relationship. Before cetaceans, raoellids are the first record of this brain mass/body mass decoupling resulting in a very low EQ and unexpected position in the ln graph illustrated in Figure 7. Counter-intuitively, the cetaceans that nowadays have the second biggest brain after humans derive from a group of mammals that had a lower-than-average expected brain size. This is probably a side effect of the adaptation to aquatic life. Available data on the volumes of the early cetaceans remain scarce [60], and a better knowledge of the brain of these taxa, like pakicetids, will allow to test this scenario further. Conversely, the very small brain size relative to body mass observed in *Khirtharia* might be another evidence supporting the aquatic habits in raoellids.

Acknowledgment
We thank Jacob Maugoust (ISEM) for fruitful discussions and Renaud Lebrun (ISEM) for the access to scanning facilities and help with scanning the material (MRI platform member of the national infrastructure France-BioImaging supported by the French National Research Agency [ANR-10-INBS-04, "Investments for the future"], the LabEx CEMEB [ANR-10-LABX-0004], and NUMEV [ANR-10-LABX-0020]).

Statement of Ethics
The material studied in this work entirely consists of fossil specimens; as such, ethics approval was not required. The fossil specimens included in this work were collected using ethical practices and are housed in public institutions.

Conflict of Interest Statement
The authors have no relevant financial or non-financial interests to disclose. The authors have no competing interests to declare that are relevant to the content of this article. All authors certify that they have no affiliations with or involvement in any organization or entity with any financial interest or non-financial interest in the subject matter or materials discussed in this manuscript.

Funding Sources
The research leading to these results received funding from the FYSSEN foundation, the Institut des Sciences de l'Evolution, Montpellier (ISEM, projet au sud), and the Royal Belgian Institute of Natural Sciences. For the purpose of Open Access, a CC-BY public copyright licence has been applied by the authors to the present document and will be applied to all subsequent versions up to the Author Accepted Manuscript arising from this submission.

Author Contributions
M.W. collected the specimens in the field. M.O. and M.W. performed the 3D segmentations and wrote the main manuscript text. M.O. prepared the figures. M.W., M.O., T.S., and R.R. reviewed and improved the manuscript.

Data Availability Statement
All the data presented in this work are available in tables and supplementary material. The 3D models deriving from the *Khirtharia* specimen will be available online for visualization and free download on the platform MorphoMuseuM (https://morphomuseum.com/). Further enquiries can be directed to the corresponding author.